\documentclass[final,3p,times, sort&compress]{elsarticle}
\usepackage{graphicx}% Include figure files
\usepackage{dcolumn}% Align table columns on decimal point
\usepackage{bm}% bold math
\usepackage{upgreek}
\usepackage{hyperref}
\usepackage{ams math}
\usepackage{multirow}
\usepackage{multicol} % Multiple columns environment
\usepackage{nomencl} % Nomenclature package

\makenomenclature

\usepackage{hyperref}
\renewcommand\nomgroup[1]{%
  \item[\bfseries
  \ifstrequal{#1}{G}{Greek symbols}{%
  \ifstrequal{#1}{S}{Subscripts}{}}%
]}

\begin{document}

\begin{frontmatter}

\title{Coalescence delay mediated by the gas layer during the impact of hot droplets}
\author{Zhigang Xu}
\author{Haicheng Qi}
\author{Tianyou Wang}
\author{Zhizhao Che\corref{cor1}}
\cortext[cor1]{Corresponding author.% Tel.: +65-67905587 Fax: +65-67911859
}
\ead{chezhizhao@tju.edu.cn}
\address{State Key Laboratory of Engines, Tianjin University, Tianjin, 300350, China.}

\begin{abstract}
Coalescence may not occur immediately when droplets impact a liquid film. Despite the prevalence of the high-temperature condition during the impact process in many applications, the effect of droplet temperature on droplet coalescence is rarely considered. In this study, we experimentally investigate the droplet coalescence during the impact of hot droplets on a liquid film by using color interferometry, high-speed imaging, and infrared imaging. We find that the coalescence of the hot droplet with the liquid film can be delayed which is mediated by the intervening gas layer between the droplet and the film. Compared with droplets at room temperature, the residence time of hot droplets can increase by more than two orders of magnitude. We find that the thickness of the gas layer increases with the droplet temperature, explaining that the thermal delay of coalescence is due to the thicker gas layer. During the hot droplet impact, the temperature gradient at the bottom of the droplet induces Maranogni flow, which can delay the drainage of the intervening gas layer. The results also show that as the Weber number increases, the residence time of the droplet decreases because of the thinner thickness of the gas layer.
\end{abstract}

\begin{keyword}
\texttt {
Droplet coalescence \sep
Coalescence delay \sep
Gas layer \sep
Temperature effect \sep
Droplet impact
}
\end{keyword}

\end{frontmatter}

\section{Introduction}\label{sec:1}
The impact of droplets on a liquid surface occurs in many natural and industrial processes, such as spray combustion \cite{Moreira2010DropletImpactICE}, spray coating \cite{Guo2016ConsiderationSurroundingAir}, spray cooling \cite{Horacek2005SingleNozzleCooling, Kim2004EvaporativeCoolingMicroporous}, and precipitation \cite{Raes2000AerosolsGlobalTroposphere}. The outcomes of a droplet impacting liquid surfaces mainly include splashing \cite{Lee2015VortexRingsSplashing, Li2018HollowDropletWetted}, jet breakup \cite{Agbaglah2015VortexSheddingJet, Michon2017JetDynamicsPost}, and bubble entrapment \cite{Thoraval2016VortexLargeBubble, Xu2018PoolTemperaturBubble}. The phenomena of droplet bouncing or partial coalescence may also occur when droplets impact a liquid film \cite{Blanchette2006, Blanchette2009DynamicsCoalescenceInterfaces}. The impact of droplets is a complex process and involves the interplay of different forces, such as the droplet inertia \cite{Leng2001SplashSphericalDrops, Riboux2014CriticalImpactSpeed}, the surface tension \cite{Deng2007ViscosityBubbleEntrapment, Zeff2000SingularityDynamicsJet}, the viscous force \cite{Langley2017UltraViscousAir, Marcotte2019EjectaCorollaSplashes}, the gravitational force \cite{Ray2010GenerationSecondaryDroplets, Yue2006CoalescenceNewtonianViscoelastic}, and the environmental pressure \cite{Li2017DoubleContactDrop, Xu2005Splashing}. Hence, the impact process is affected by many parameters, such as the diameter of the droplet, the impact speed of the droplet, and the properties of the droplet (e.g., density, viscosity, and surface tension).

Upon the impact of droplets on a liquid surface, the bouncing of the droplet is possible \cite{Tang2019BouncingDropFilm} owing to the presence of a thin gas layer between the droplet and liquid surface before the droplet wets the liquid surface. Even though the intervening gas layer is very thin, it can significantly affect the impact dynamics. With the interferometry technique, many details of the gas layer under different impact parameters have been unveiled \cite{Bouwhuis2012MaximalAirBubble, Lo2017AtomicallySmoothSubstrate}. The gas layer usually has the shape of a dimple, collapses at the edge of the gas layer, and leads to the entrapment of a bubble beneath the droplet \cite{Qi2020AirHeatedSubstrates, Veen2012AirLayerProfiles}. The gas layer can also collapse at many points during the impact of the ultra-viscous droplet \cite{Langley2017UltraViscousAir}, which is different from the one-point collapse of the gas layer for the water droplet. At reduced environmental pressure, the thickness of the gas layer decreases \cite{Li2017DoubleContactDrop}, and the center thickness of the gas layer decreases by about 85.7$\%$ when the environmental pressure decreases from 19.7 kPa to 2.8 kPa. The substrate temperature can affect the thickness of the gas layer \cite{Qi2020AirHeatedSubstrates}, and the average thickness of the gas layer on the heated substrate at 70 $^\circ$C is about 12$\%$ thicker than the droplet on the unheated substrate.

The effect of temperature on droplet coalescence was rarely studied, even though high/low-temperature conditions are ubiquitous in applications. Studies have shown that two droplets of different temperatures can repel each other without coalescence when they get closer \cite{Dell1996SuppressionCoalescenceTemperature, Dell1998LiquidsStayDry}. Savino et al.\ \cite{Savino2003MarangoniFlotationDroplets} found that the coalescence of the droplet on a liquid pool could be delayed by increasing the temperature of the liquid pool due to the Marangoni motion. Michela et al.\ \cite{Geri2017ThermalDelayCoalescence} found that the delay time of the droplet increased with the temperature difference between the cold droplet and the hot pool following a 2/3-power law. Mogilevskiy found that a cold droplet could levitate and move on a hot liquid pool surface \cite{Mogilevskiy2020LevitationNonboilingDroplet}.

The previous studies have shown the importance of temperature on droplet coalescence, but it is still unknown how a hot droplet impacts a liquid film, even though it often exists in many applications, such as the fuel atomization in internal combustion engines \cite{Panao2005SprayImpingementInjection} and the dispersed flow in nuclear power plants \cite{Xie2004AnnularMistFlow, Zhong2019HeatMistFlow}. In this study, we find that during the impact of a hot droplet on a cold liquid film, the coalescence is delayed which is mediated by the gas layer between the droplet and the liquid film. The delayed coalescence is because of the Marangoni flow induced by the temperature gradient at the bottom of the droplet, which resists the drainage of the gas layer. Through high-speed imaging and color interferometry, we unveil the detailed evolution of the gas layer.

\section{Experimental setup}\label{sec:2}

\begin{figure*}
  \centering
  \includegraphics[scale=0.6]{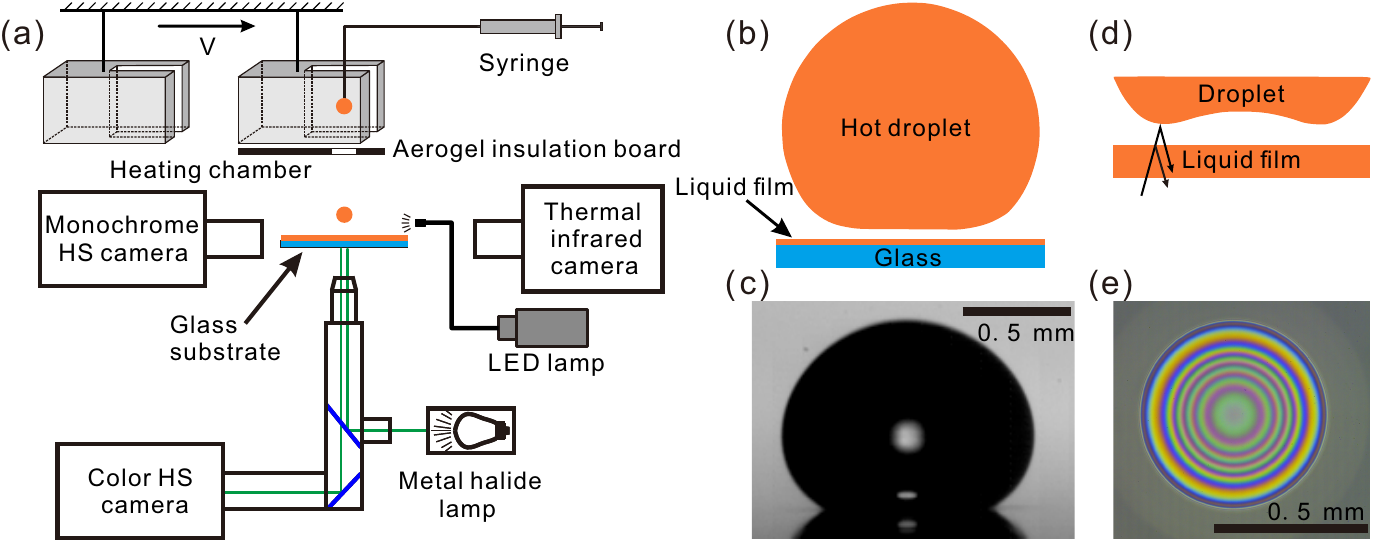}
  \caption{(a) Schematic diagram of the experimental setup. (b) Schematic diagram of a hot droplet impacting a liquid film. (c) A typical image from the side view. (d) Interference schematic diagram of the intervening gas layer between the droplet and the liquid film. (e) A typical image of the interference fringes from the bottom view.
}\label{fig:01}
\end{figure*}

The experimental setup for a hot droplet impacting a liquid film is illustrated in Fig.\ \ref{fig:01}. Droplets were released from the tip of a syringe needle, which was connected to a syringe pushed by a syringe pump (Harvard Apparatus, Pump 11 elite Pico plus), to impact the liquid film. The droplet and the liquid film are silicone oil with 200 cSt. The droplet was pre-heated by thermal radiation in a through slot of the cubic heating block. Heating coils were embedded in the walls of the heating block and a proportional-integral-derivative (PID) temperature controller was used to control the temperature of the heating block. After each experiment, the heating block was moved away to avoid the continuous heating of the liquid. An aerogel insulation board with a round hole was fixed underneath the heating block to avoid heating the liquid film, while the heated droplet can fall through the hole. The actual temperature of the droplet was obtained from the images captured by an infrared camera (INFRATEC ImageIR 8355BBhp) at a resolution of 272 $\times$ 512 pixels, corresponding to a spatial resolution of 19.8 $\mu$m/pixel. The temperature range of the droplet in this study is below 114.2 $^\circ$C, and the vaporization of the silicone oil droplet in the experiment is negligible. The liquid film was coated on a glass slide laying on the horizontal microscope stage. The thickness of the liquid film was 100 $\mu$m, which was controlled by its volume. The schematic diagram of a hot droplet impacting a liquid film is shown in Fig.\ \ref{fig:01}b. The speed of the droplet impact was varied by changing the height of the syringe needle from the surface of the liquid film, and the actual speed of the droplet was measured from the side-view images via image processing using a customized Matlab program. The process of droplet impact was captured simultaneously by two high-speed cameras from the side view and the bottom view. The side-view images were captured by a high-speed monochromatic camera (Photron Fastcam SA-1.1) with microlens (Nikon ED AF Micro Nikkor 200 mm f/4D), and the processes of droplets impact were illuminated by a high-power LED lamp (Hecho S5000).

\begin{figure*}
  \centering
  \includegraphics[scale=0.32]{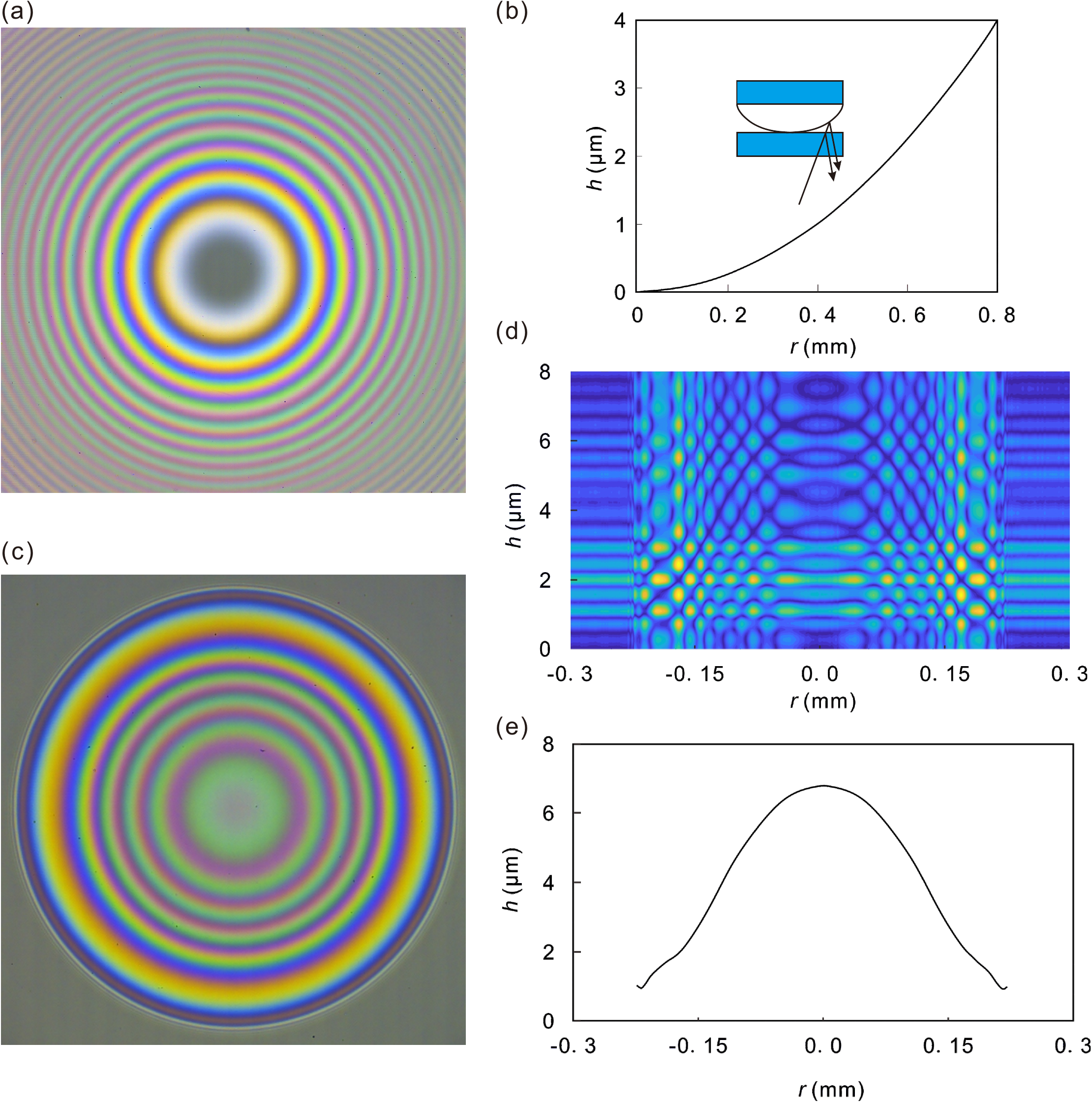}
  \caption{Method to obtain the thickness of the gas layer from the interference images. (a) Interference fringes of the plano-convex lens. (b) Thickness of the gas layer underneath the plano-convex lens. Inset: schematic of the setup used to obtain the interference fringes of the gas layer. (c) Interference fringes of the gas layer during the droplet impact. (d) Color-difference map by calculating the Euclidean distance \cite{Qi2020AirHeatedSubstrates, Veen2012AirLayerProfiles}. (e) Thickness profile of the gas layer.
}\label{fig:02}
\end{figure*}

To obtain the high-speed evolution of the gas layer between the droplet and the liquid film, the color interference measurement was performed during the impact process. The interference fringes from the bottom-view images were captured by a high-speed color camera (Photron Fastcam SA-1.1) with an inverted microscope (Nikon Ti-U with $10\times $ objective lens). The white light emitted from a metal halide lamp (Nikon intense light C-HGFI) was used as an interference source through an optical fiber. The green lines in Fig.\ \ref{fig:01}a represent the optical path in the interference measurement. The optical path was integrated into the microscope by a translucent semi-reflective mirror and a reflective mirror. The interference schematic of the gas layer between the droplet and the liquid film is described in Fig.\ \ref{fig:01}d. A typical interference image is shown in Fig.\ \ref{fig:01}e. The side-view image corresponding to the interference image is shown in Fig.\ \ref{fig:01}c. The two high-speed cameras were all at the speed of 5000 frames per second (fps) with a resolution of 1024 $\times$ 1024 pixels.

The thickness of the gas layer was obtained from the bottom-view images via image processing following the procedure in Ref.\ \cite{Qi2020AirHeatedSubstrates, Veen2012AirLayerProfiles}, which is briefly described here. A plano-convex lens with a known radius of curvature (129.24 mm) was used for the calibration, and its interference fringes are shown in Fig.\ \ref{fig:02}a, and the thickness of the gas layer underneath the plano-convex lens is shown in Fig.\ \ref{fig:02}b. The interference fringes of a typical gas layer when the droplet impact are shown in Fig.\ \ref{fig:02}c. Fig.\ \ref{fig:02}d shows the color-difference map, which was generated by calculating the Euclidean distance \cite{Qi2020AirHeatedSubstrates, Veen2012AirLayerProfiles} between the color profiles in Lab color spaces obtained from the interference fringes of the plano-convex lens (Fig.\ \ref{fig:02}a) and gas layer (Fig.\ \ref{fig:02}c). Finally, the thickness of the gas layer can be obtained from the color-difference map, as shown in Fig.\ \ref{fig:02}e. Please refer to Ref.\ \cite{Qi2020AirHeatedSubstrates, Veen2012AirLayerProfiles} for more details about the method.

\begin{figure*}
  \centering
  \includegraphics[scale=0.6]{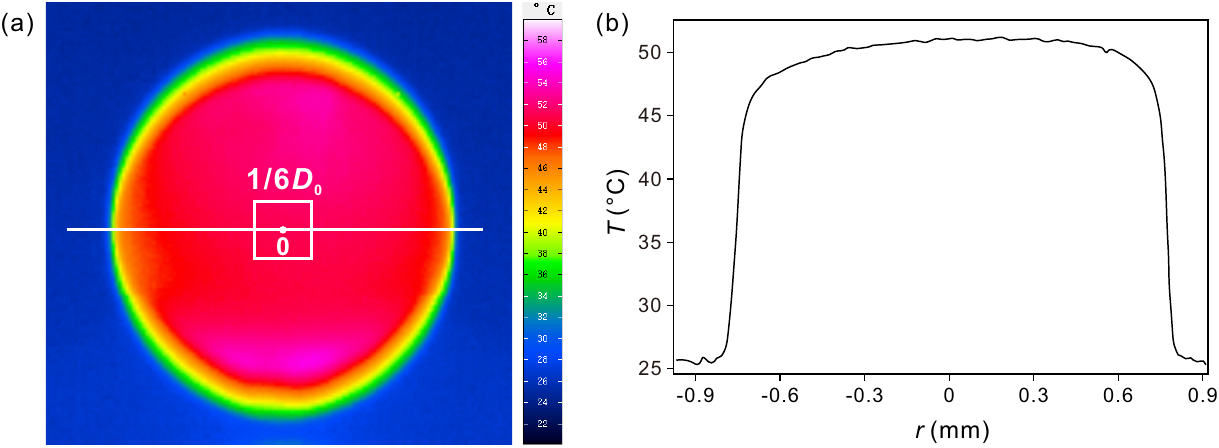}
  \caption{Method to determine the temperature of the hot droplet. (a) Temperature distribution of the droplet surface. The white line is a radial line passing through the center point. The square at the center of the droplet with the size of $1/6D_0 \times 1/6D_0$ is used to calculate the mean temperature of the droplet. (b) Temperature distribution along the white line in Fig.\ \ref{fig:03}a.
}\label{fig:03}
\end{figure*}

The droplet temperature was measured from the infrared image just before the impact. The typical temperature distribution of the droplet surface captured by the infrared camera is shown in Fig.\ \ref{fig:03}a. The vertical edge of the droplet is blurry because of the vertical movement of the droplet during the exposure interval. A thermocouple was used to calibrate the droplet temperature captured by the infrared camera, and the emittance in the central square area of the droplet surface is 0.96 based on the calibration data. Fig.\ \ref{fig:03}b shows the temperature distribution along the white line in Fig.\ \ref{fig:03}a. The temperature of the hot droplet is obtained by calculating the average temperature of a square region of the droplet surface ($1/6D_0 \times 1/6D_0 $ in size at the center of the droplet, indicated by the white box in Fig.\ \ref{fig:03}a) using a customized Matlab program.

\section{Results and discussions}\label{sec:3}
\subsection{Thermal delay of coalescence in the droplet impact}\label{sec:31}
\begin{figure*}
  \centering
  \includegraphics[scale=0.55]{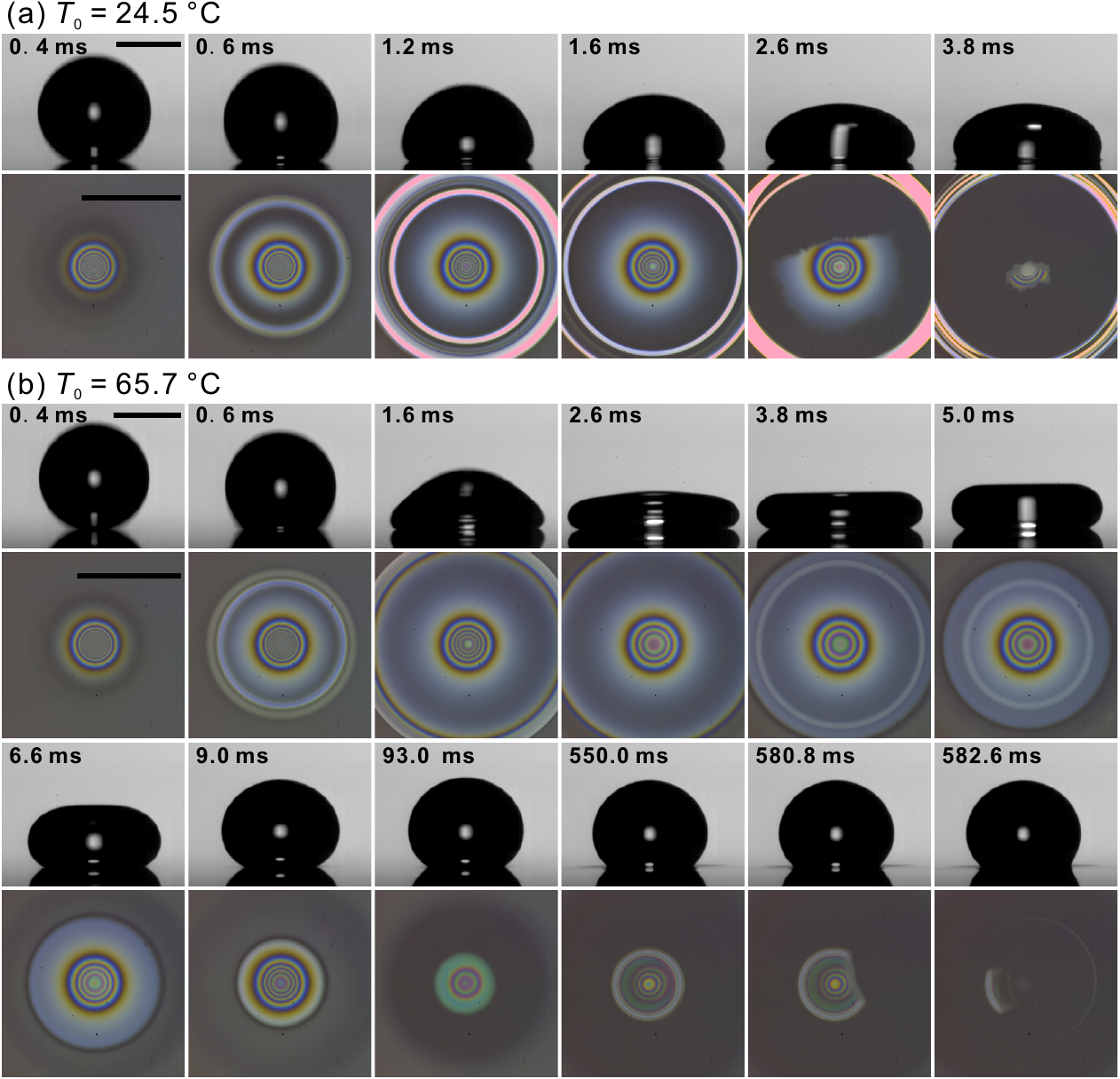}
  \caption{Impact processes of two droplets of different temperatures on a liquid film ($D_0 = 1.59$ mm and $U_0 = 0.63$ m/s). (a) Droplet at room temperature, $T_0 = 24.5$ $^\circ$C (Supplementary Movie 1). (b) Droplet with a high temperature, $T_0 = 65.7$ $^\circ$C (Supplementary Movie 2). The first rows are the side-view images and the second rows are the interference fringes captured from the bottom view. The scale bars are 1 mm.
}\label{fig:04}
\end{figure*}

To consider the effect of droplet temperature on the impact dynamics, we used droplets of different temperatures to impact a liquid film, and the results show that the coalescence of the hot droplet with the liquid film is significantly delayed in comparison with the droplet at the room temperature. The impact processes for two typical droplets are shown in Fig.\ \ref{fig:04}, among them, one is a hot droplet (65.7 $^\circ$C, Fig.\ \ref{fig:04}b), and the other is at room temperature (24.5 $^\circ$C, Fig.\ \ref{fig:04}a). Here we define the moment when the first clear interference fringe is seen from the bottom-view images as $t = 0$. The first rows in Figs.\ \ref{fig:04}a and b are droplet morphology from the side-view images and the second rows are the interference fringes from the bottom-view images. We can observe that the coalescence for the impact of the hot droplet is significantly delayed. For the droplet at room temperature, the droplet can stay on the liquid film for only about 2 ms before the droplet coalescence, but for the hot droplet, the droplet can stay on the liquid film for about 580 ms, which is two orders of magnitude longer than that at room temperature. The delayed coalescence of the hot droplet is mediated by the intervening gas layer between the droplet and the liquid film, whose evolution can be seen from the interference fringes in the bottom view images. When the droplet is at room temperature, the gas layer beneath the droplet is squeezed to be wide when the droplet lands, and sparse interference fringes are formed at the outer region of the gas layer, as shown at 0.6, 1.2, and 1.6 ms in Fig.\ \ref{fig:04}a, indicating that the outer region of the gas layer is relatively flat but the inner region is steeper, consisting with previous studies \cite{Qi2020AirHeatedSubstrates, Veen2012AirLayerProfiles}. The interference fringes finally collapse when the droplet wets the liquid film at 2.6 ms. In contrast, for the hot droplet, the droplet retracts and rebounds (at $t > 2.6$ ms) after it spreads to the maximum, and the diameter of the gas layer decreases during the droplet retraction. After that, the droplet can stay steadily on the liquid film for a long time, as shown in Fig.\ \ref{fig:04}b from 9.0 to 550.0 ms. Finally, the droplet wets the liquid film and the interference fringes collapse at 580.8 ms.

\subsection{Effect of droplet temperature}\label{sec:32}

\begin{figure}
  \centering
  \includegraphics[scale=0.38]{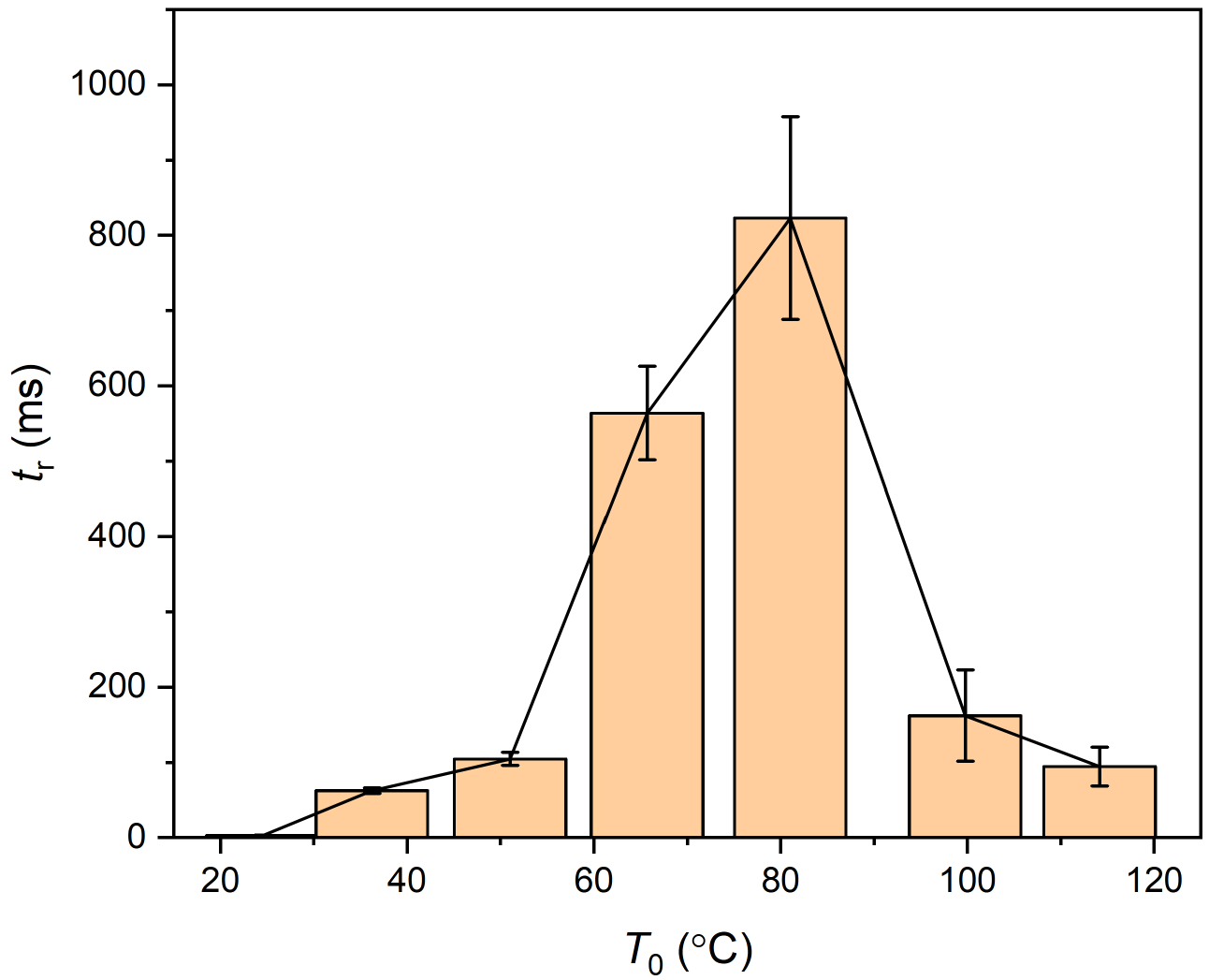}
  \caption{Effects of droplet temperature on the residence time of the droplet. $D_0 = 1.59$ mm and $U_0 = 0.63$ m/s. The error bars indicate the standard deviations of repeated experiments of more than 8 times.
}\label{fig:05}
\end{figure}

To analyze the delayed coalescence of the hot droplet mediated by the gas layer, we varied the droplet temperature in a wide range and focused on the droplet residence time and the gas layer. The effects of droplet temperature on the residence time of the droplet are shown in Fig.\ \ref{fig:05}. The residence time of the droplet $t_r$ is defined as the time interval from the first instant that clear interference fringe can be seen ($t = 0$) in the bottom-view images to the instant that the gas layer starts to rupture. As the droplet temperature increases from room temperature (24.5 $^\circ$C) to 114.2 $^\circ$C, the residence time of the droplet first increases and then decreases. When the droplet temperature is 24.5 $^\circ$C, the residence time of the droplet on the liquid film is about only 2 ms. When the droplet temperature increases to 81.0 $^\circ$C, the residence time of the droplet increases to 820 ms, which is about 400 times longer than that at room temperature. However, beyond that, the trend of residence time reverses. When the droplet temperature further increases to 114.2 $^\circ$C, the residence time of the droplet decreases to 95 ms.

\begin{figure*}
  \centering
  \includegraphics[scale=0.55]{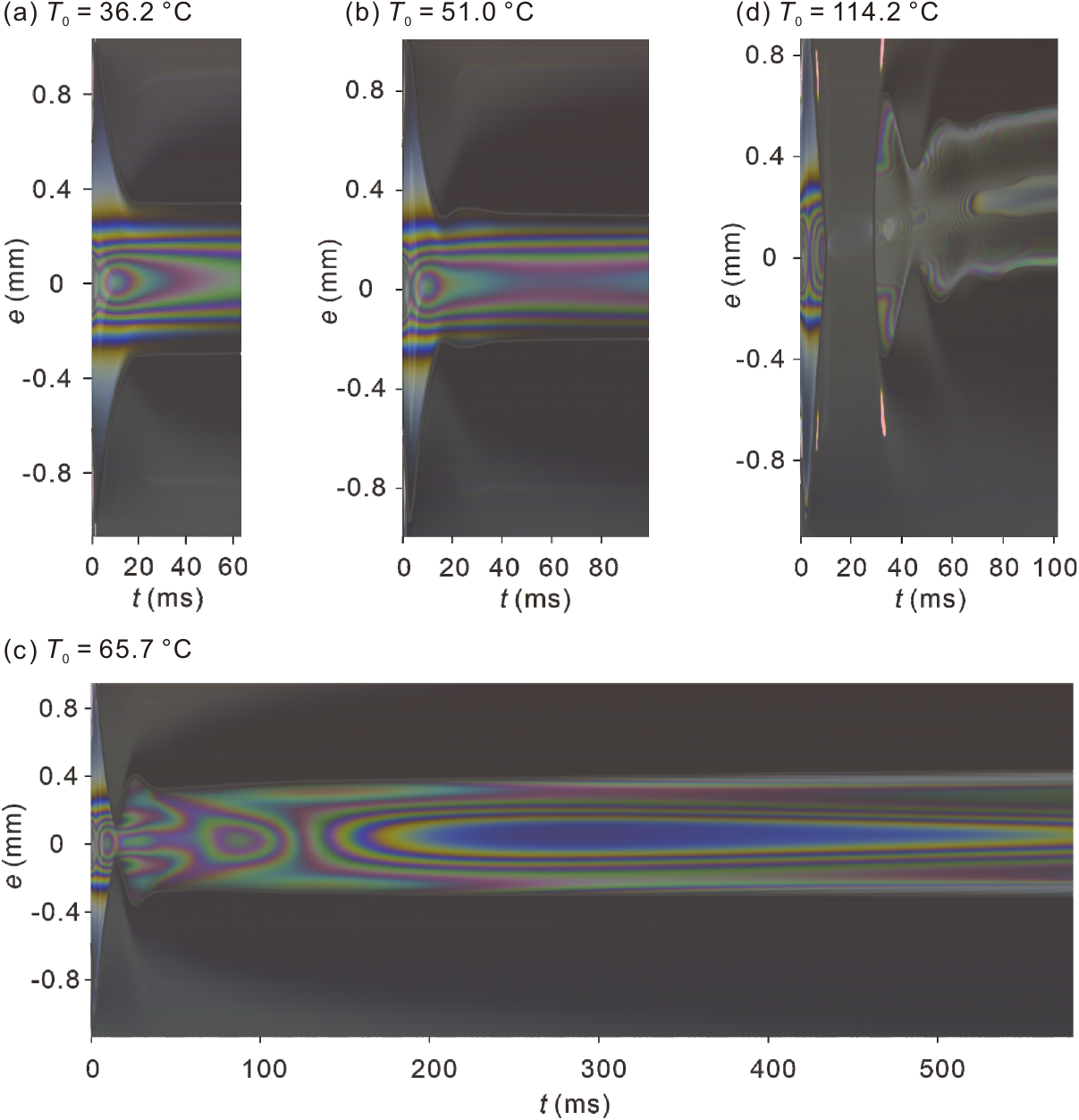}
  \caption{Time-space images of the interference fringes showing the entire processes of the droplets impacting a liquid film. (a) $T_0 = 36.2$ $^\circ$C. (b) $T_0 = 51.0$ $^\circ$C. (c) $T_0 = 65.7$ $^\circ$C. (d) $T_0 = 114.2$ $^\circ$C. $D_0 = 1.59$ mm and $U_0 = 0.63$ m/s. The images are obtained by stacking the interference fringes along a cross-section at different time steps.
}\label{fig:06}
\end{figure*}

The residence time of the droplet is directly correlated to the evolution of the intervening gas layer. To show the entire process of the gas layer evolution, we plot time-space images of the interference fringes, as shown in Fig.\ \ref{fig:06}, which were obtained by stacking the interference fringes along a cross-section at different time steps. Hence, each vertical line in the time-space image corresponds to the interference fringes along the cross-section at a specific instant, and the whole time-space image can show the entire process of the interference fringes. When the droplet temperature is 36.2 $^\circ$C (see Fig.\ \ref{fig:06}a), the width of the interference fringes (the distance between the upper and lower boundaries of the interference fringes, which is a measure of the size of the gas layer) is almost a constant ($t > 20$ ms) as the droplets rest steadily on the liquid film surface. When the droplet temperature increases to 65.7 $^\circ$C (see Fig.\ \ref{fig:06}c), there is a stage at about 15 ms when the width of the interference fringes increases and then decreases. This is because the droplet vibrates slightly upon the impact, which induces the variation of the interference fringes correspondingly. When the droplet temperature is very high (114.2 $^\circ$C, see Fig.\ \ref{fig:06}d), the interference fringes disappear from 15 to 30 ms because the hot droplet rebounds upon the impact and leaves the liquid film surface. We can also see that the droplet deviates from the initial impact point after the rebound, indicating a slight migration of the droplet. Meanwhile, the gas layer is asymmetric during the migration of the droplet, indicating the development of instability at the bottom of the droplet (probably the Kelvin-Helmholtz instability \cite{Liu2015KelvinHelmholtzInstability} induced by the airflow in the gas layer and the Marangoni flow on the interface). During the whole process, the thickness profile of the gas layer changes quickly and greatly, and the bottom of the droplet is easier to contact with the surface of the liquid film, thus causing the residence time of the droplet on the liquid film to decrease (see Fig.\ \ref{fig:05}). Later in this section, we mainly focus on the temperature range below 81.0 $^\circ$C, because above this range of droplet temperature, the droplet may migrate and the process is less controllable.

\begin{figure*}
  \centering
  \includegraphics[width=\columnwidth]{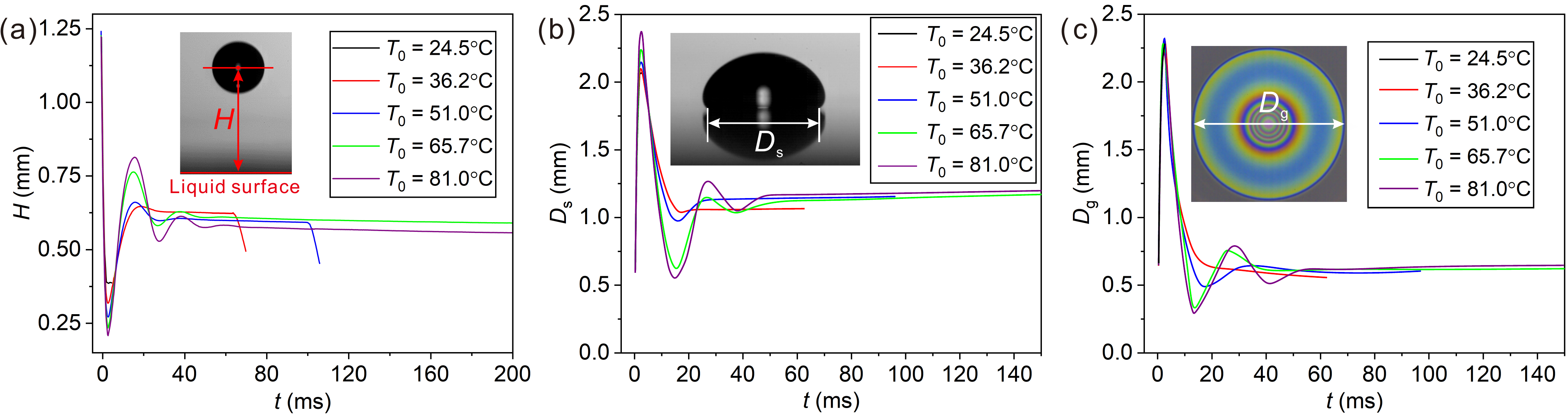}
  \caption{Effects of droplet temperature on (a) the height of the droplet \emph{H}, (b) the spreading diameter of the droplet $D_s$, and (c) the diameter of the gas layer $D_g$. $D_0 = 1.59$ mm and $U_0 = 0.63$ m/s.
}\label{fig:07}
\end{figure*}

To quantitatively analyze the evolution of the droplet and the gas layer during the impact of hot droplets, the height of the droplet, the spreading diameter of the droplet, and the diameter of the gas layer are obtained from the high-speed images via digital image processing. The height of the droplet \emph{H} is measured from the centroid of the droplet to the surface of the liquid film in the side-view images (see the inset illustration in Fig.\ \ref{fig:07}a), the spreading diameter of the droplet $D_s$ is measured from the outmost boundary at the bottom of the droplet in the side-view images (see the inset illustration in Fig.\ \ref{fig:07}b), and the diameter of the gas layer $D_g$ is measured from the maximum diameter of the interference fringes of the gas layer in the bottom-view images (see the inset illustration in Fig.\ \ref{fig:07}c). For the droplet at the room temperature, as the droplet falls and spreads due to the inertia, the droplet height decreases, but the spreading diameter of the droplet and the diameter of the gas layer increases quickly. Then the droplet contacts the liquid film. When the droplet temperature is 36.2 $^\circ$C, the droplet height decreases quickly and then increases as the droplet retracts after it spreads, meanwhile the diameter of the gas layer increases and then decreases. Then the droplet height decreases slowly as the droplet falls slowly due to gravity, reaching a quasi-equilibrium state. Finally, the droplet height decreases suddenly after 64 ms because the intervening gas layer ruptures and the droplet wets the liquid film. As the droplet temperature increases further to 65.7 and 81.0 $^\circ$C, the curves of the droplet height fluctuate several times as the droplet vibrates, as shown in Fig.\ \ref{fig:07}a. From Figs.\ \ref{fig:07}b and c, we can see that the diameter of the gas layer is almost constant as the droplet reaches the quasi-equilibrium state (\emph{t} $\textgreater$ 50 ms), and the spreading diameter of the droplet is slightly larger than the diameter of the gas layer because of the difference in the resolution between the side-view images and the bottom-view interference fringes. By comparing the curves at different temperatures, we can see that the droplet temperature can remarkably affect the evolution of the droplet height and the diameter of the gas layer. As the droplet temperature increases, the minimum of the droplet height during the droplet oscillation (at about 3 ms) decreases and occurs as the droplet spreads to the maximum extent, as shown in Fig.\ \ref{fig:07}a. Besides, as the droplet temperature increases, the maximum of the droplet height (at about 15 ms) increases and occurs as the droplet retracts to the maximum extent. As the droplet temperature increases, the minimum diameter of the gas layer during the droplet retraction (from about 10 to 25 ms) decreases, as shown in Fig.\ \ref{fig:07}c. This is because the surface tension and the viscosity of the liquid decrease as the temperature increases, and the droplet of lower surface tension and the viscosity deforms more easily. Therefore, as the droplet temperature increases, the droplet jumps higher because of the larger retraction (see 10 - 25 ms in Fig.\ \ref{fig:07}a), and it results in a small diameter of the gas layer.

\begin{figure*}
  \centering
  \includegraphics[scale=0.25]{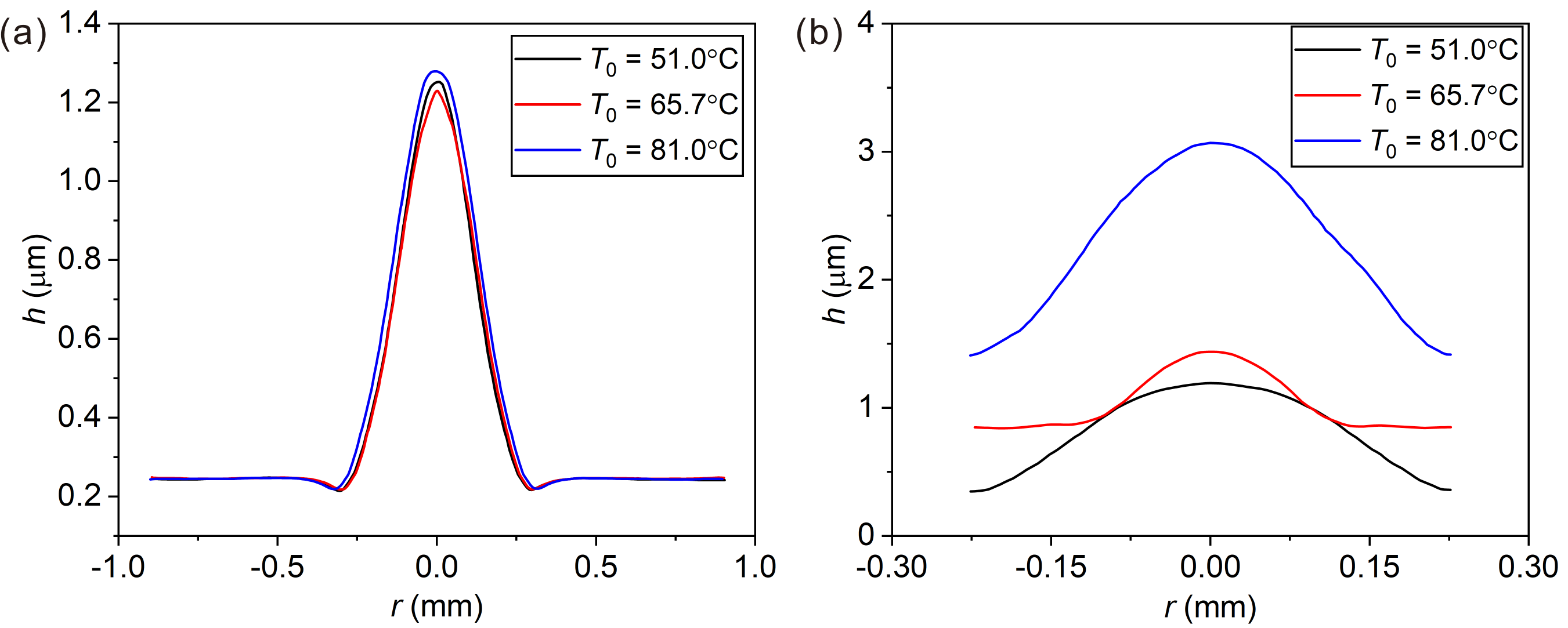}
  \caption{Profiles of the gas layer thicknesses at two instants after the droplets of different temperatures impact a liquid film. (a) $t = 2$ ms. (b) $t = 21.6$ ms. $D_0 = 1.59$ mm and $U_0 = 0.63$ m/s.
}\label{fig:08}
\end{figure*}

Before the hot droplet wets the liquid film, the air underneath the droplet is compressed to form a gas layer of a dimple shape. The shape of the gas layer plays an important role in the residence time of the droplet. We extracted the thickness profile of the gas layer from the interference images, and the typical profiles of the gas layer thicknesses at two instants after the impact are plotted in Fig.\ \ref{fig:08}. Immediately upon the impact, the profiles of the gas layer thickness at different droplet temperatures are almost the same, as shown in Fig.\ \ref{fig:08}a. This is because the droplet inertia dominates at the initial stage of the impact and the effect of the temperature is negligible. Hence, the gas layers are squeezed by the same droplet inertia and deformed to the same extent at the bottom of the droplets.

\begin{figure*}
  \centering
  \includegraphics[scale=0.7]{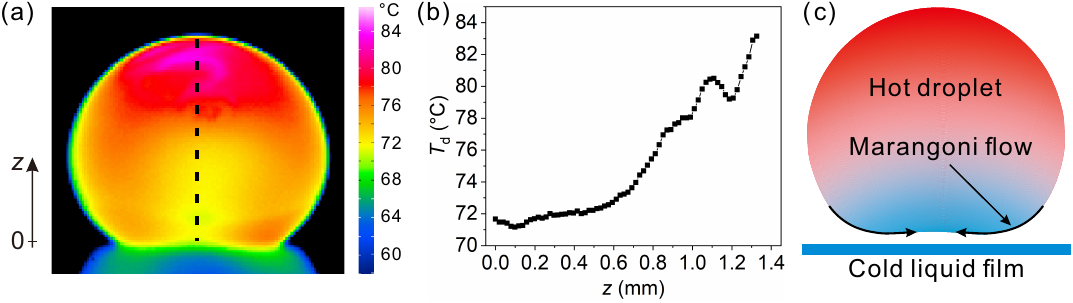}
  \caption{(a) Typical infrared image of a hot droplet surface. $D_0 = 1.59$ mm, $U_0 = 0.63$ m/s, $T_0 = 81.0$ $^\circ$C, and \emph{t} = 117.2 ms. (b) Temperature distribution along the black dotted line in Fig.\ \ref{fig:09}a. (c) Schematic diagram of the Marangoni flow for a hot droplet impacting a cold liquid film. The Marangoni flow at the bottom of the droplet can resist the drainage of air from the gas layer or even entrain air into the gas layer between the droplet and the liquid film.
}\label{fig:09}
\end{figure*}

In the later stage, the difference in the gas layer profile becomes significant. As the droplet temperature increases, the gas layer at $t = 21.6$ ms becomes much thicker, as shown in Fig.\ \ref{fig:08}b. This could be explained using the Marangoni effect, which is induced by the temperature difference between the droplet and the liquid film \cite{Aversana1996CoalescenceWettingGas, Dell1998LiquidsStayDry}. The time scale for the heat transfer at the bottom of the droplet can be estimated by considering the heat conduction in the gas layer, and it is about $10^{-8}$ s \cite{Qi2020AirHeatedSubstrates}. In this stage, the droplet inertia is small as the droplet has reached the quasi-equilibrium shape and stays steadily on the liquid film. Meanwhile, because of the heat transfer from the hot droplet to the cool liquid film, the local temperature at the bottom of the droplet is lower than the main part of the droplet, as shown in Fig.\ \ref{fig:09}a, a typical infrared image of a hot droplet surface. The temperature distribution along the black dotted line in Fig.\ \ref{fig:09}a is shown in Fig.\ \ref{fig:09}b, which also shows that the temperature of the droplet surface increases from the bottom to the top of the droplet. Hence, the temperature difference can induce the variation of the surface tension and then Marangoni flow along the surface of the droplet, as shown in Fig.\ \ref{fig:09}c. Therefore, the Marangoni flow at the bottom of the droplet can resist the drainage of air from the gas layer. As the droplet temperature increases, the Marangoni effect becomes stronger and the surface flow at the bottom of the droplet becomes faster, which can slow down the drainage of the air or even entrain air into the gas layer between the droplet and the liquid film. Therefore, as the droplet temperature increases, the gas layer becomes thicker when the droplet stays steadily on the liquid film, as shown in Fig.\ \ref{fig:08}b.

\begin{figure*}
  \centering
  \includegraphics[width=\columnwidth]{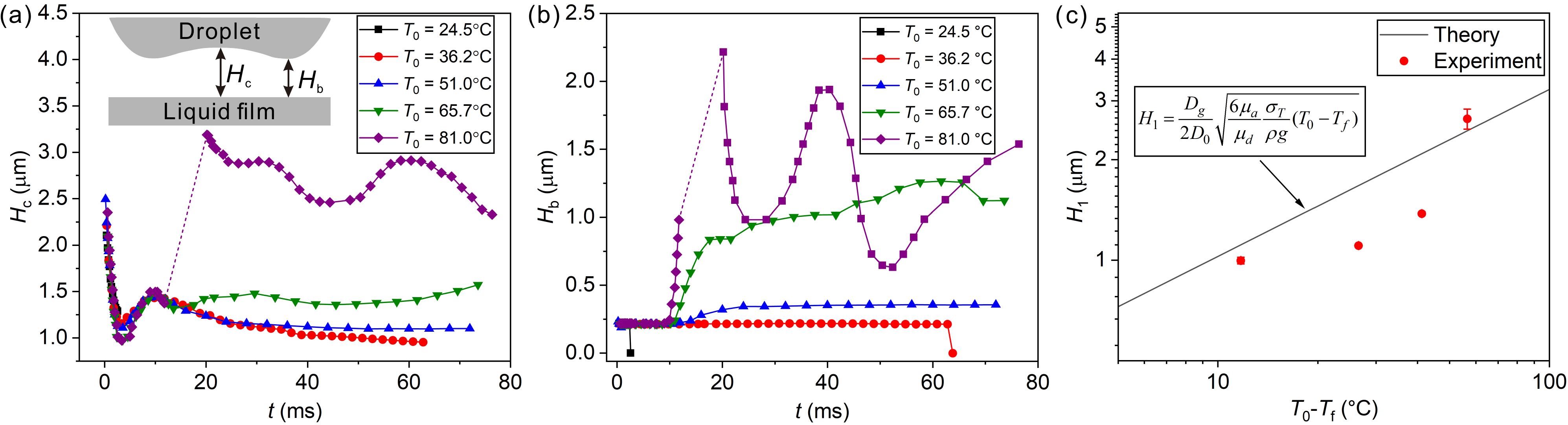}
  \caption{Gas layer thickness for the impact of the droplets of different temperatures on a liquid film. (a) Thickness of the gas layer at the center point. (b) Thickness of the gas layer at the kink. (c) Comparison of the gas layer thickness between the experiment and the model by Eq.\ (\ref{eq:4}). $D_0 = 1.59$ mm and $U_0 = 0.63$ m/s. For $T_0 = 81.0$ $^\circ$C, the dashed line at 12 -- 20 ms indicates that the droplet rebounds, the interference fringes disappear, and the film thickness cannot be obtained.
}\label{fig:10}
\end{figure*}

To further quantify the evolution of the gas layer thickness during the impact process, the thickness of the gas layer at the center point $H_c$ and at the kink $H_b$ (see the schematic in Fig.\ \ref{fig:10}a) is selected to characterize the thickness of the gas layer. The time evolutions of $H_c$ and $H_b$ are shown in Figs.\ \ref{fig:10}a and b. For the droplet at room temperature (24.5 $^\circ$C), $H_c$ decreases quickly initially as the droplet lands on the liquid film under gravity, as shown in Fig.\ \ref{fig:10}a. $H_b$ quickly goes to zero as the droplet coalescence with the liquid film, as shown in Fig.\ \ref{fig:10}b. When the droplet temperature is 36.2 $^\circ$C, $H_c$ decreases quickly initially as the droplet lands on the liquid film, then increases as the droplet shape vibrates. Finally, $H_c$ decreases slowly when the droplet reaches the quasi-equilibrium state, as shown in Fig.\ \ref{fig:10}a. In contrast, $H_b$ is almost unchanged, as shown in Fig.\ \ref{fig:10}b, because the gas at the center point of the gas layer flows outward to replenish the gas layer under the kink. Finally, $H_b$ quickly goes to zero as the droplet coalescence with the liquid film. Because $H_b$ is greatly affected by $H_c$ (see 20 -- 70 ms at $T_0$ = 65.7 and 81.0 $^\circ$C in Fig.\ \ref{fig:10}b) and vibrates violently, we mainly discuss the variation of gas layer thickness at the center point $H_c$ with droplet temperature $T_0$. When the droplet temperature is 81.0 $^\circ$C, there is a sudden jump in the curve of $H_c$ at $12 < t <20$ ms. This is because of the bouncing of the droplet on the liquid film, hence the gas layer is so thick that the interference fringes cannot form. Then the droplet falls and impacts the liquid film again at about 20 ms because of the gravity. The curve of $H_c$ oscillates several times as the droplet vibrates on the liquid film surface.

Comparing the thickness evolution of the gas layer at different droplet temperatures, $H_c$ in the initial stage of the droplet impact ($t < 3$ ms) almost does not vary with the droplet temperature because the droplet inertia dominates. But later in the quasi-equilibrium stage ($t > 20$ ms), the higher droplet temperature results in a thicker gas layer. This is because the droplet inertia is small in this stage, but the heat transfer from the hot droplet to the liquid film reduces the local temperature at the bottom of the droplet. Hence the temperature gradient induces Marangoni flow along the droplet surface as illustrated in Fig.\ \ref{fig:09}c. As the droplet temperature increases, the Marangoni flow becomes stronger, which can slow down the drainage of the gas layer or even entrain air into the air film. Therefore, the gas layer at the quasi-equilibrium stage becomes thicker as the droplet temperature increases. The thicker gas layer can delay the droplet from contacting the liquid film, and increases the residence of the droplet.

The thickness of the gas layer in the quasi-equilibrium stage can be considered using the lubrication theorem. Because of the small thickness of the gas layer, the lubrication flow in the gas layer induces a pressure increment in the gas layer \cite{Savino2003MarangoniFlotationDroplets}:
\begin{equation}\label{eq:1}
  \Delta P=\frac{{{\mu }_{a}}{{U}_{m}}}{H_{1}^{2}}{{D}_{0}},
\end{equation}
where ${{\mu }_{a}}$ is the viscosity of the air, $H_1$ is the thickness of the gas layer, and $U_m$ is the thermal characteristic Marangoni speed, which can be calculated as \cite{Savino2003MarangoniFlotationDroplets}:
\begin{equation}\label{eq:2}
  {{U}_{m}}={{\sigma }_{T}}({{T}_{0}}-{{T}_{f}})/{{\mu }_{d}},
\end{equation}
where ${{\sigma }_{T}}$ is the surface tension temperature coefficient, ${{\mu }_{d}}$ is the viscosity of the droplet, and ${{T}_{f}}$ is the surface temperature of the liquid film.

Because of the quasi-equilibrium state of the droplet, the inertia of the droplet is negligible and the lubricating pressure balances the weight of the droplet
\begin{equation}\label{eq:3}
  \frac{1}{4}\Delta P\pi {D_g}^{2}=\frac{1}{6}\pi \rho gD_{0}^{3},
\end{equation}
where $D_g$ is the diameter of the gas layer. Substituting Eqs.\ (\ref{eq:1}) and (\ref{eq:2}) into Eq.\ (\ref{eq:3}), the thickness of the gas layer $H_1$ can be expressed as
\begin{equation}\label{eq:4}
  {{H}_{1}}=\frac{{{D}_{g}}}{{2{D}_{0}}}\sqrt{\frac{6{{\mu }_{a}}}{{{\mu }_{d}}}\frac{{{\sigma }_{T}}}{\rho g}({{T}_{0}}-{{T}_{f}})}.
\end{equation}
From Eq.\ (\ref{eq:4}), we can get that the gas layer thickness increases with the droplet temperature. To further check the above analysis, we compare Eq.\ (\ref{eq:4}) with experimental data. We can use ${{D}_{g}}\sim 0.6$ mm because the diameter of the gas layer $D_g$ is almost constant when the droplet is in the quasi-equilibrium state, as shown in Fig.\ \ref{fig:07}c. By substituting the experimental data ($D_0$ = 1.59 mm, $T_f$ = 24.5 $^\circ$C, \emph{g} = 9.8 m$/$s$^2$, $\mu_a$ = 0.01809 mPa$\cdot$s, $\mu_d$ = 193.4 mPa$\cdot$s, $\sigma_T$ = 5 $\times$ $10^{-5}$N$/$m$\cdot$$^\circ$C, and $\rho$ = 967 kg$/$m$^3$) into Eq.\ (\ref{eq:4}), the theoretical relationship between the gas layer thickness and the droplet temperature is plotted in Fig.\ \ref{fig:10}c (see the solid line). The experimental results are obtained by calculating the average values of $H_c$ between 40 and 60 ms (relatively stable stage) in Fig.\ \ref{fig:10}a, and included in Fig.\ \ref{fig:10}c (see the symbols) for comparison. The error bars indicate the standard deviations of $H_c$ between 40 and 60 ms in Fig.\ \ref{fig:10}a. The comparison shows that the experimental results agree reasonably well with the theoretical results.

It should be noted that the evolution of thickness of the gas layer $H_c$ is not always synchronous with the evolution of the droplet height $H$. For example, when the droplet temperature is 81.0 $^\circ$C, the droplet height $H$ increases at 3 -- 15 ms because the droplet retracts and leaves the liquid film surface, as shown in Fig.\ \ref{fig:07}a. Meanwhile, $H_c$ increases at 3 -- 10 ms, as shown in Fig.\ \ref{fig:10}a, and has the same trend of variation with the droplet height $H$. However, after that, $H_c$ suddenly decreases at 10 -- 12 ms. This is because the capillary waves propagate along the droplet surface towards the center of the droplet during the droplet retraction, and converge at the center of the droplet bottom, making the bottom interface of the droplet move downward. Therefore, $H_c$ suddenly decreases at 10 -- 12 ms, as shown in Fig.\ \ref{fig:10}a, even though $H$ increases.

\begin{figure}
  \centering
  \includegraphics[width=0.45\columnwidth]{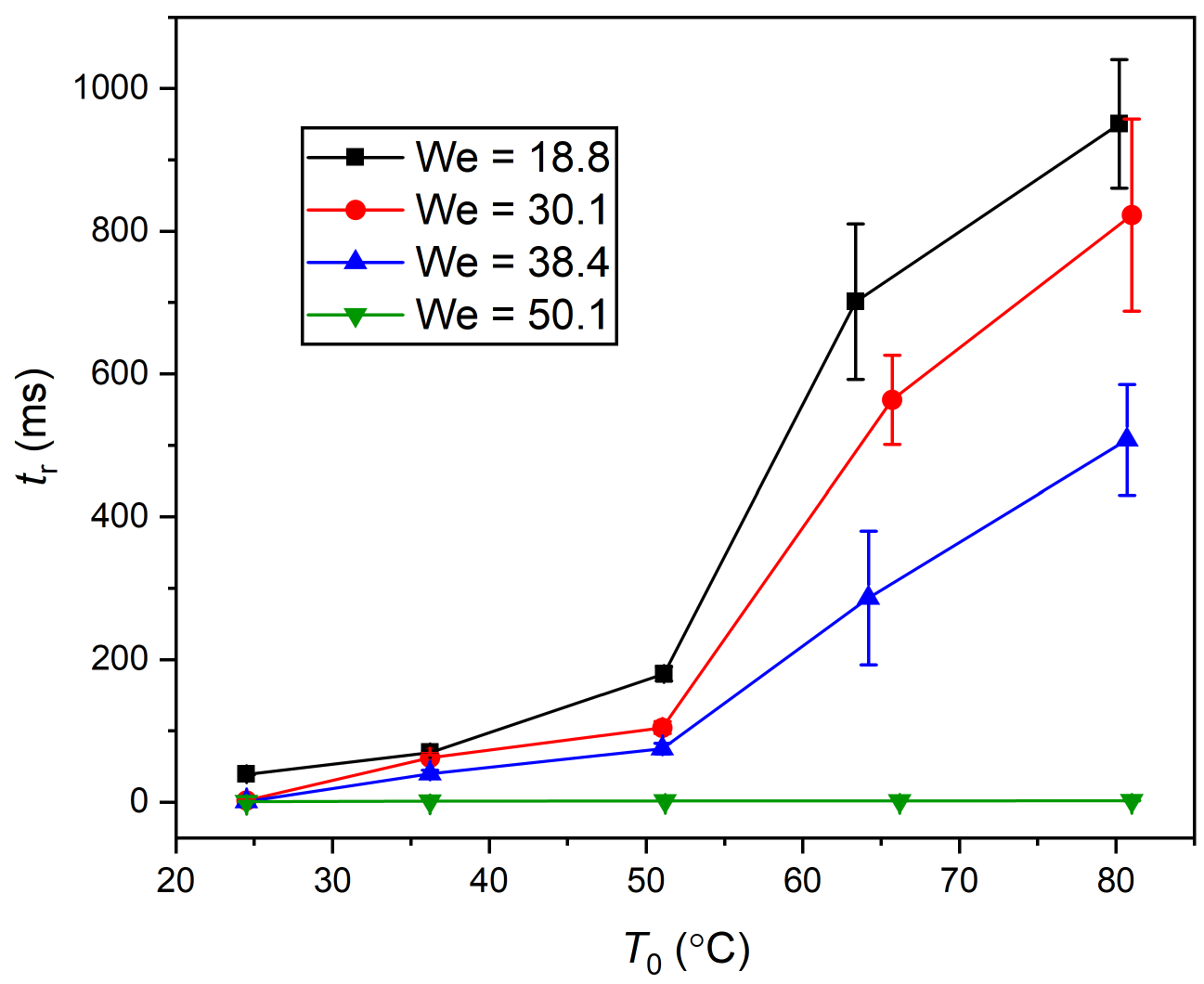}
  \caption{Effects of the Weber number on the residence time of the droplet. $D_0 = 1.59$ mm. The error bars indicate the standard deviations of repeated experiments of more than 8 times.
}\label{fig:11}
\end{figure}

\subsection{Effect of droplet inertia}\label{sec:33}
The droplet Weber number $\text{We}\equiv \rho U_{0}^{2}{{D}_{0}}/\sigma $ can be used to measure the ratio between the droplet inertia and the capillary force, where $\rho$ and $\sigma$ are the density and surface tension of the liquid. To analyze the effect of the droplet inertia on the coalescence, we varied the droplet Weber number from 18.8 to 50.1 by varying the impact speed of the droplet in the experiments. Fig.\ \ref{fig:11} shows the effects of the inertia and the temperature of the droplet on the residence time of the droplet. The residence time of the droplet decreases as the Weber number increases. At a higher Weber number (We = 50.1), the residence time of the droplet is very short because the droplet almost immediately wets the liquid film as the droplet with higher inertia can quickly overcome the lubrication pressure in the gas layer. In addition, the residence time of the droplet increases with the droplet temperature (because of the increase in the gas layer thickness as discussed in Section \ref{sec:32}), and this variation tendency is almost the same for droplets of different Weber numbers. The dependence of the residence time on the Weber number could be attributed to the gas layer, which will be discussed next.

\begin{figure*}
  \centering
  \includegraphics[width=\columnwidth]{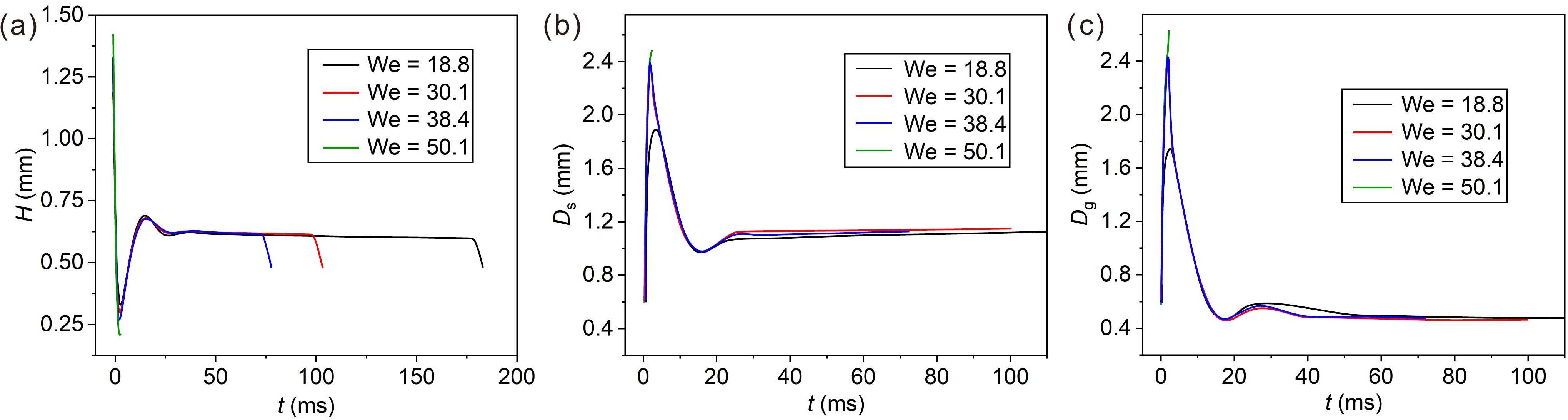}
  \caption{Effects of Weber number on (a) the height of the droplet \emph{H}, (b) the spreading diameter of the droplet $D_s$, and (c) the diameter of the gas layer $D_g$. $D_0 = 1.59$ mm and $T_0 = 51.0$ $^\circ$C.
}\label{fig:12}
\end{figure*}

To quantify the evolution of the droplet shape at different Weber numbers, we first consider the variations of the droplet height, the spreading diameter of the droplet, and the diameter of the gas layer at different Weber numbers, as shown in Fig.\ \ref{fig:12}. At a higher Weber number (We = 50.1), the droplet height decreases, but the spreading diameter of the droplet and the diameter of the gas layer increase quickly as the droplet falls and spreads due to the inertia. Then the droplet contacts the liquid film soon. At lower Weber numbers (We = 18.8, 30.1, and 38.4), the droplet height decreases quickly and then increases as the droplet retracts after it spreads, and the spreading diameter of the droplet and the diameter of the gas layer increases and then decrease. Then the droplet height decreases slowly as the droplet enters the quasi-equilibrium stage. The spreading diameter of the droplet and diameter of the gas layer are almost constant when the droplet is in the quasi-equilibrium stage at $t > 50$ ms. As the Weber number increases, the minimum height of the droplet (at about 3 ms) decreases, and the maximum diameter of the gas layer increases. This is because, as the droplet inertia increases, the droplet deforms more severely after the impact.

\begin{figure*}
  \centering
  \includegraphics[width=0.8\columnwidth]{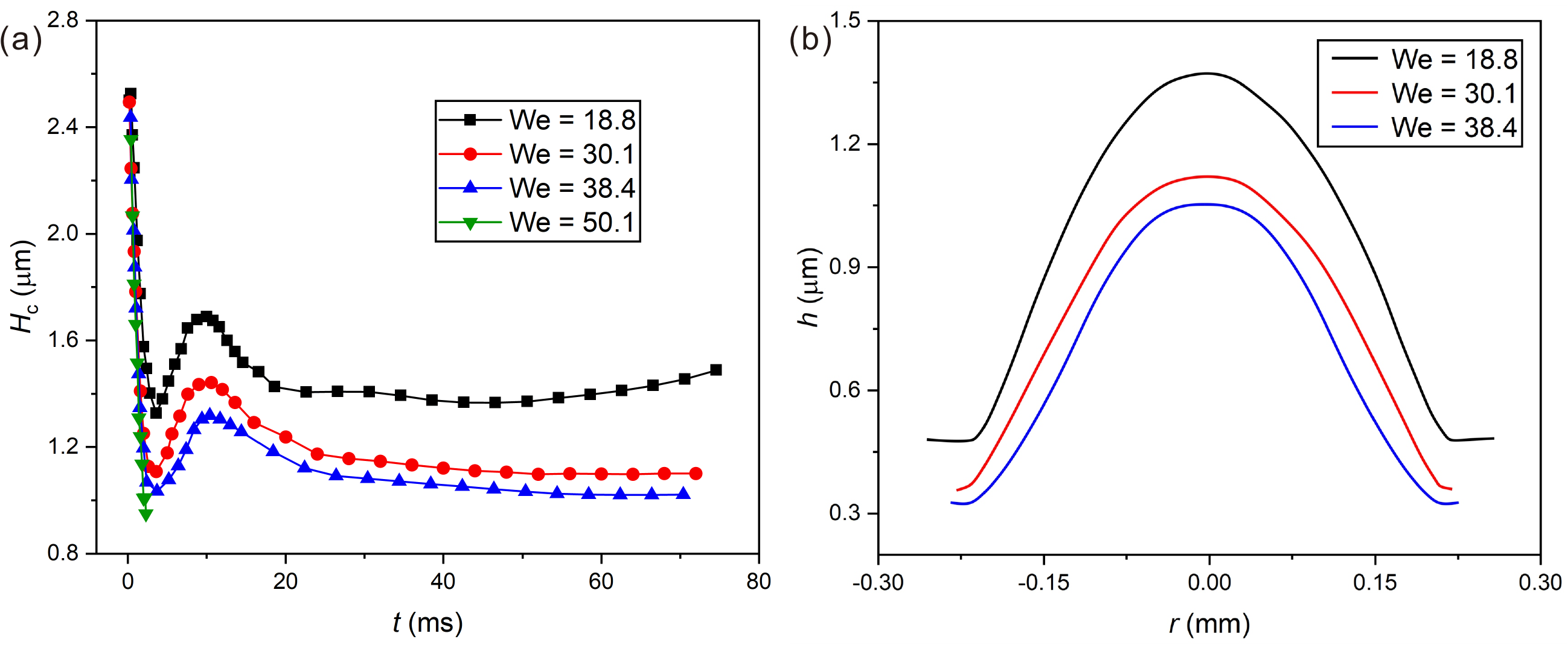}
  \caption{Effects of Weber number on the gas layer for the impact of hot droplets on a liquid film. (a) Evolution of the gas layer thickness. (b) Profiles of the gas layer thicknesses at $t = 40$ ms. $D_0 = 1.59$ mm and $T_0 = 51.0 ^\circ$C.
}\label{fig:13}
\end{figure*}

To consider the effect of the Weber number on the gas layer, the evolution of the gas layer thickness at the center point $H_c$ is compared for different Weber numbers of the droplet impact, as shown in Fig.\ \ref{fig:13}a. At We = 50.1, $H_c$ decreases quickly as the droplet falls and spreads due to the droplet inertia. Then the droplet contacts the liquid film immediately. At We = 30.1 and 38.4, $H_c$ decreases quickly and then increases, and finally decreases slowly in the quasi-equilibrium stage. At We = 18.8, $H_c$ first decreases and then increases at $t > 30$ ms because the droplet vibrates on the liquid film surface. The typical profiles of the gas layer thicknesses in the quasi-equilibrium stage at different Weber numbers are plotted in Fig.\ \ref{fig:13}b. The thickness of the gas layer decreases as the Weber number increases, which facilitates the droplet from contacting the liquid film. Therefore, the residence time of the droplet decreases as the Weber number increases, as discussed in Fig.\ \ref{fig:11}.

\section{Conclusions}\label{sec:4}
In this paper, we experimentally study the impact of hot droplets on liquid films, and found that the coalescence of the hot droplet with the liquid film can be delayed due to the intervening gas layer between the droplet and the film. As the droplet temperature increases from the room temperature to 81.0 $^\circ$C, the residence time of the droplet increases by more than two orders of magnitude. Further increasing the droplet temperature will lead to the reduction in the residence time of the droplet because of the instability development in the gas layer. By using color interferometry and high-speed photography, the evolution of the intervening gas layer between the droplet and the liquid film is obtained. We find that the thickness of the gas layer increases with increasing the droplet temperature, explaining that the thermal delay of coalescence during the impact of the hot droplet is due to the thicker gas layer. The thicker gas layer is due to the Marangoni flow induced by the temperature gradient at the bottom of the droplet, which resists the drainage of the gas layer. We also study the effect of the Weber number on the coalescence of the droplet and the evolution of the gas layer. The results show that as the Weber number increases, the residence time of the droplet decreases because the gas layer becomes thinner.

Through this experimental study, we have shown that the coalescence of the hot droplet with the liquid film can be delayed due to the intervening gas layer. There are still many open questions about the process, such as the theoretical analysis of the thermal delay of coalescence during the impact of the hot droplets, the mechanism of the rupture of the gas layer, and the subsequent dynamics of the wrapped bubble produced after the rupture of the gas layer.

\section*{Declaration of Competing Interest}
None.

\section*{Acknowledgements}
This work is supported by the National Natural Science Foundation of China (Grant Nos.\ 51676137, 52176083 and 51921004).

\section*{Supplementary material}
Supplementary material associated with this article can be found in the online version.

\bibliography{CoalescenceDelay}

\end{document}